\documentclass[english,prb,showpacs,amsmath,amssymb,aps,12pt]{revtex4-1}

\usepackage{amsfonts}
\usepackage{amssymb}
\input epsf
\usepackage[utf8]{inputenc}
\usepackage[english]{babel}
\usepackage{esint}
\usepackage{braket}
\usepackage{graphicx}
\usepackage{epstopdf}
\usepackage{pdfsync}
\usepackage[dvipsnames]{xcolor}
\usepackage{hyperref}
\usepackage{slashed}
\usepackage{dsfont}
\usepackage[normalem]{ulem}
\usepackage{cancel}

\newcommand{\be}[1]{ \begin{eqnarray} \mbox{$\label{#1}$} }
\newcommand{\ee}{\end{eqnarray}}
\newcommand{\eeq}{\end{equation}}

\newcommand\grad{\mathbf\nabla}

\newcommand\half{\frac 1 2 }

\newcommand{\av}[1]{\langle #1\rangle}

\newcommand\noi{\noindent}

\newcommand\oncite [1] {Ref.\,\onlinecite{#1} }

\begin{document}
\title{Fractionalized anyons in counterflowing Quantum Hall Liquids}

\author{Jun-Xiao Hui}
\affiliation{Tsung-Dao Lee Institute, Shanghai Jiao Tong University, Shanghai, 201210, China}
\author{T.H. Hansson}
\affiliation{Department of Physics, Stockholm University, AlbaNova University Center, 106 91 Stockholm, Sweden }
\author{Egor Babaev}
\affiliation{Department of Physics, KTH-Royal Institute of Technology, SE-10691, Stockholm, Sweden}
\affiliation{Wallenberg Initiative Materials Science for Sustainability, Department of Physics, KTH Royal Institute of Technology, SE-106 91 Stockholm, Sweden}

\begin{abstract}
A key property of topologically ordered systems, such as Quantum Hall states, is the existence of excitations obeying fractional quantum statistics---anyons.
We develop a theory for multicomponent counterflow states where an ordinary Laughlin quasiparticle can split into fractional vortices carrying fractions of its charge and statistical angle. There are two phases, separated by a quantum phase transition, where in the first, although observable, the fractionalized charges are asymptotically confined. In the second phase, they are unconfined anyons and the topological order is different from that of the Laughlin state.
\end{abstract}

\maketitle

\noindent
\emph{\bf{Introduction}}:
A plethora of condensed matter phases can be understood in terms of Bose condensates, either of elementary fields like in $^4$He, or of Cooper pairs in $^3$He and electronic superconductors.   Helium atoms, although composite, can at large distances be described by a local field. This description also extends to topologically ordered states where many quantum Hall (QH) liquids can be described as Bose condensates of composite bosons\cite{ZHK1989,zhang1992chern,10.21468/SciPostPhys.8.5.079}, and Cooper pairs of composite fermions\cite{greiter1992paired,moore1991nonabelions}. In the simplest cases, the condensation is of a single-component field, but the generalization to many components introduces new phenomena. Examples include superconducting vortices carrying arbitrary fractions of the flux quantum \cite{Babaev2002vortices,Iguchi2023,zheng2024observation,Zhou2024}. Examples of multicomponent topological states are quantum Hall liquids with new filling fractions\cite{girvin1996multicomponent}. A fundamental characteristic of all these states is various kinds of superflow of combinations of charge, mass, spin etc. A particular kind of flow in a two-component system is counter-flow, where two components flow in the opposite direction with the same speed, so that the total density is unchanged, while mass and charge can flow if the quantum numbers and masses differ between the components. A trivial example of such counter-flow appears in condensates of excitons, which is a flow of Coulomb-bound states of an electron and a hole, where the flow carries momentum but no charge. A more subtle example is the edge currents in quantum spin Hall liquids\cite{kane2005quantum} which are not related to any bound state.

Qualitatively new symmetry-breaking states arise in multicomponent systems when counter-flows involve correlated motion of pairs of electron pairs\cite{babaev2002phase,babaev2004superconductor}, or pairs of bosons \cite{kuklov2003counterflow,kuklov2004superfluid}.
Such flows are related to ``higher-order" condensates, i.e., states where various four- or six-fermion operators have non-zero expectation values. Importantly, these states cannot be described in terms of bound states of many electrons. An illustrative example is the Borromean supercounterfluid in three-component systems where all components can flow, but none of them has a fixed, and not even defined, type of counter-flowing ``partner", in contrast to condensates of locally bound states of two or more particles \cite{blomquist2021borromean,babaev2024hydrodynamics,babaev2024topological,kuklov2026field}.

So far, most of the proposed examples of ``higher-degree" counter-flow were non-topological multicomponent symmetry-breaking states and composed of either exclusively fermionic or exclusively bosonic mixtures. In this paper, we go beyond the most investigated cases and study counter-flow states that are topologically ordered and have both fermionic and bosonic components. Similarly to a fractional quantum Hall liquid, the topological order in our system originates from a strong magnetic field, but the resulting \emph{counter-flowing Quantum Hall liquids} (CFQH) have distinct properties that differ from usual QH states.

These properties emanate from the fact that unit vortices in such multicomponent systems can fractionalize in an arbitrary way. The energetics of the fractionalization is similar to the recently experimentally observed fractionalization of single-quantum vortices in multicomponent superconductors \cite{Iguchi2023,Zhou2024,zheng2024observation}, but the physical interpretation and physical consequences are different. In the CFQH liquids, the vortices, and thus the quasiparticle charges, are fractionalized. There are two phases, the first with the same topological order as a Laughlin liquid, where the quasiparticles can have arbitrary charges. Although these objects are stable and physically relevant, asymptotically, they are logarithmically confined. In a second phase, which has a distinct topological order, the quasiparticle charges are quantized, but differently from the Laughlin case, and we have unconfined anyons. In the specific three component model discussed next, their electric charge is $e^\star = e/2k$ and the fractional statistics angle $\theta = \pi/2k$.

\noindent
\emph{\bf{The model}}:
We take a mixture of two neutral bosons, $\phi_{n}$, $n=1,2$ with masses $m_i$, and one unit charge fermion $\psi$ with mass $m_f$. We assume a strong background magnetic field $B$, and for simplicity, and we shall further assume that the chemical potentials are such that $\av{\rho_f} = \av{\rho_1} = \av{\rho_2} = \bar\rho$.
Most importantly, we assume that we are in a Mott regime  where the total density is frozen, and equals $\bar\rho$ with no fluctuations, which also means that the total particle current vanishes. One way to impose the Mott constraint is to make intercomponent coupling slightly weaker than intracomponent coupling in a Hubbard model\cite{Kuklov2004,kuklov2003counterflow,soyler2009sign,altman2003phase,Sellin18superfluiddrag,blomquist2021borromean}. However, here we shall just assume the constraint. An important difference between these works and this paper is that we have counter-flows between bosons and fermions, which are fermionic and carry electric charge (or synthetic electric charge, like in e.g. the case of optical lattice emulators \cite{galitski2019artificial,Dalibard}).

Next, we consider a background magnetic field $B$ sufficiently strong for the fermionic mode to form a FQH state at a Laughlin filling factor $\nu=1/k$.
Importantly, the fermionic mode is composite, so transport of the original fermions is possible only with a backflow, i.e., exchange of positions of either of the two bosons;  in a more general setting, such counterflows have been referred to as ``Borromean"  \cite{blomquist2021borromean,babaev2024hydrodynamics,golic2025borromean,kuklov2026field}. This, as we see below, will have important consequences for the topological properties of the model.

Assuming the external parameters (temperature, impurity density, magnetic field $B$) are in the QH regime, we
use the Ginzburg-Landau-Chern-Simons construction\cite{ZHK1989} and first turn the fermion into a boson, $\psi \rightarrow \phi_f$, by coupling it to a statistical gauge potential $a$ with the Chern-Simons action,
\be{csact}
S_{CS} = \frac 1 {4\pi k} \int d^3x\, \epsilon^{\mu\nu\sigma} a_\mu \partial_\nu a_\sigma
\ee
where $k$ is an odd integer. For a moment consider the case where we ignore the Mott constraint for total density; then, for the bosons we take the Lagrangian 
\be{bosact1}
{\cal L}
&=& -\rho_f (\partial_0\theta_f + a_0 ) - \frac {\rho_f } {2m_f} (\vec\nabla \theta_f - \vec a {\color{orange}-} e\vec A)^2  \\
&+& \sum_{n=1}^2 \left[-\rho_n (\partial_0\theta_n - \mu_n) - \frac {\rho_n } {2m_b} (\vec\nabla \theta_n -\vec C )^2  \right]  - V[\rho_f,\rho_1,\rho_2] \nonumber
\ee
where $\vec A$ includes both the strong constant background magnetic field $\vec B = \vec\nabla\times \vec A_{b}$, and a small probe field $\delta \vec A$ used to calculate electric charges and currents. We also introduced an auxiliary probe gauge field $\vec C$ which will be used to extract the total bosonic edge current; note that it couples to both the bosonic fields. In writing \eqref{bosact1}, we used a polar representation $\phi_{f,1,2} = \sqrt \rho_{f,1,2} e^{i\theta_{f,1,2}}$, neglected the spatial derivatives of the density, and assumed that the repulsive potential $V$ and the boson chemical potentials $\mu_n$ are  chosen  to minimize the potential energy at the filling factor $\nu=1/k$ for the fermionic component.

Using the notation $\theta_\pm = \half (\theta_1 \pm \theta_2),\, \rho_{\pm} = \rho_1 \pm \rho_2$ and denoting deviations of the boson densities from the spatially constant mean values by $\delta\rho_\pm = \rho_\pm - \av{\rho_\pm} ,\, \delta\rho_f = \rho_f-\av{\rho_f}$, the Lagrangian \eqref{bosact1} becomes,
\be{bosact2}
{\cal L} &=& - \delta\rho_f \partial_0 \theta_f - a_0 \rho_f - \frac {\bar\rho}{2m_f} (\vec\grad \theta_f - \vec a - e\vec A)^2
 -\delta \rho_+ \partial_0 \theta_+  - \delta\rho_- \partial_0 \theta_- - \frac {\bar\rho}{m_b} (\vec\grad \theta_+ -\vec C)^2 \nonumber  \\
 &-&   \frac {\bar\rho}{m_b} (\vec\grad \theta_-)^2    - V[\rho_f,\rho_1,\rho_2]  + \frac 1 {4\pi k}   \epsilon^{\mu\nu\sigma} a_\mu \partial_\nu a_\sigma  \, .
\ee
$a_0$ is a multiplier field that ties the fermionic density to the statistical magnetic field $b=\epsilon^{ij}\partial_i a_j$, and in the mean field approximation where $\vec a = -e\vec A$, we get the $2\pi k \bar\rho_f = eB $, which is the Streda relation. We also only kept the leading order in the density fluctuations.

Using the Mott constraint on the total density,
$
\delta\rho_f +  \delta\rho_+ = 0 \, ,
$
the kinetic part of \eqref{bosact2} becomes
$
-\delta\rho_f(\partial_0\theta_f- \partial_0\theta_+) - \delta\rho_- \partial_0 \theta_-  \equiv -\delta\rho_f \partial_0\tilde \theta  - \delta\rho_- \partial_0 \theta_-  \, ,
$
where $\tilde\theta  = \theta_f - \half \theta_+$.
The zero current constraint,
$
m_b (\vec\nabla \theta_f - \vec a - e\vec A) +  m_f (\vec\nabla \theta_+ -\vec C)=0
$
follows from the Mott constraint, and that the individual currents are conserved. It can be used to eliminate $\vec\nabla\theta_+$, and for the special case $m_f = m_b$, we get the Lagrangian,\\

\be{bosact3}
{\cal L} &=& -\delta \rho_f \partial_0 \tilde \theta  -  a_0 \rho_f - \frac {\bar\rho}{2\mu} (\vec\grad \tilde\theta  - \delta\vec a + e\vec \delta A + \vec C )^2    - \delta\rho_- \partial_0 \theta_-  -  \frac {\bar\rho}{m_b} (\vec\grad \theta_-)^2 \\
 &-& V[\rho_f,\rho_1,\rho_2]  + \frac 1 {4\pi k}   \epsilon^{\mu\nu\sigma} a_\mu \partial_\nu a_\sigma   \, . \nonumber
\ee
where $\mu=3m/2$, and and where we added the probe field $\vec \delta A$ that will be used to extract the electromagnetic current.

The basis $(\tilde\theta, \theta_-)$ was chosen to decouple the charged $\tilde\theta$ mode from the  neutral $\theta_-$ modes. The physics is, however, independent of the choice of basis and of the assumption of equal masses, so no fine-tuning is needed. Also for the case $m_1=m_2\neq m_f$ there is a neutral mode, but for general masses and/or densities, all modes are charged, see supp. material.

Importantly, the QH response implied by \eqref{bosact3} is that of the usual $\nu=1/k$ Laughlin liquid, but we will see that other observables, the edge modes and the quasiparticles, differ from those of the Laughlin state; keeping in mind the crucial importance of the microscopically dictated $2\pi$-periodicity of the original phase variables, we now proceed to analyze these quantities.

\vskip 2mm
\noi
\emph{\bf{Fractional vortices and fractionalization of anyons:}}
In the GLCS theory, vortices describe quasiparticles, and the vorticity is directly related to the electric charge by $2\pi k\delta\rho = \delta\Phi$. Thus classifying the vortices is important for determining the topological properties of the CFQH liquids.

We start with a three-component model, hence vortices can be classified by three integer winding numbers $(N_f,N_1,N_2)$, of the phase variables $\theta_f,\theta_1,\theta_2 \in (0...2\pi]$ i.e. $ \oint \nabla \theta_1 =2\pi N_1$. The two phase variables $\tilde\theta$ and $\theta_-$ are invariant under the compact gauge transformation $\theta_\alpha \rightarrow \theta_\alpha + \phi$, $\alpha = f,1,2$, where $\phi$ is time independent but allows for $2\pi$ jumps. As explained in \cite{babaev2024hydrodynamics}, this implies a modular identification of vortices: $(N_f,N_1,N_2)$ and $(N_f+N,N_1+N,N_2+N)$ cannot be distinguished. Using this we now show that there are three fundamental vortices:
i) The winding $(1,0,0)$ gives $\tilde\theta$ a $2\pi$ winding but leaves $\theta_-$ unchanged. Because of the coupling to the gauge potential $\vec {A} $, this is a charged, finite energy, unit vortex with a core.  ii) The windings $(0,1,0)$ or $(0,0,1)$ give $\tilde\theta$ a $-\pi$ winding, corresponding to a \emph{charged half vortex}, but also a $\pm \pi$ winding in $\theta_-$, giving a logarithmically diverging energy $\propto\frac{\bar\rho}{m_b}\ln(L/\xi_{(0,1,0)})$, or $\propto\frac{\bar\rho}{m_b}\ln(L/\xi_{(0,0,1)})$  where $L$ is the size of the system and $\xi$ is a cutoff length associated with vortex core. In general, this size can be different for different components, but we will neglect the difference here. 
iii) Combining two elementary half vortices with opposite fermionic currents i.e. taking $(0,1,-1)$, leaves $\tilde\theta$ unchanged, but amounts to a $2\pi$ winding of $\theta_-$. This vortex is purely bosonic and again has a logarithmically diverging energy. 

Because of the Chern-Simons flux-charge relation, fractionalization of vorticity implies a fractionalization of the quasiparticle charge beyond that in the Laughlin states. For the above case, the minimal fractional charge is $e/2k$, but for arbitrary masses and fillings, you generally get arbitrary fractional charges, just as magnetic flux is arbitrary fractionalized in two-component superconductors \cite{Babaev2002vortices}. Concretely, for nonsymmetric components, $m_f \neq m_1 \neq m_2$ in general, the Hamiltonian for our model can be rewritten as
 
\begin{eqnarray}
\label{eq:3comphamiltonianmottv2}
H =\int d^2x\, \Biggl[  \frac{\gamma_f(\gamma_1+\gamma_2)}{2\gamma_{tot}} \Biggl(\vec\grad\tilde\vartheta - \vec a - e\vec A \Biggr)^2 
+\frac{\gamma_1\gamma_2}{2(\gamma_1+\gamma_2)}(\vec\grad\theta_1-\vec\grad\theta_2)^2 \Biggr]
\end{eqnarray}
where $\tilde\vartheta =\theta_f-\frac{\gamma_1}{\gamma_1+\gamma_2}\theta_1-\frac{\gamma_2}{\gamma_1+\gamma_2}\theta_2$,
$\gamma_a=\frac{\bar\rho_a}{m_a}$, and $\gamma_{tot}=\gamma_f+\gamma_1+\gamma_2$.
Equation \ref{eq:3comphamiltonianmottv2} implies that the $(0,1,0)$ vortex carries a $\frac{\gamma_1}{\gamma_1+\gamma_2}$ fraction of flux and the $(0,0,1)$ vortex the fraction $\frac{\gamma_2}{\gamma_1+\gamma_2}$. A technical comment: the variables $\tilde\theta$ and $\theta_-$ are not independent, since the commutation relations with the densities have cross terms, this is, however, not of relevance for the below discussion, see the supp. material for details.   

The energy penalty due to the non-zero gradient term in $\theta_-$ implies that the vortices $(0,1,0)$ and $(0,0,1)$ are logarithmically confined and a finite size pair of such vortices will at long distances look like a fermionic $(0,1,1)$ vortex, which is identical to the $(-1,0,0)$. In spite of the confinement, these vortices can, once created, be stable and could be excited, just as their observed superconducting counterparts \cite{Iguchi2023,zheng2024observation,Zhou2024}. Also, just as fractional vortices in superconductors, they could become stable near boundaries even though energy divergence is linear in the bulk \cite{silaev2011stable,maiani2022vortex}. If so, they would not just be expensive bulk excitations, but in fact become the dominant excitations near boundaries. 

It has been shown that in the analogous case of superconductors with $U(1)^N$ symmetry, a vortex with a flux fraction different from 1 or 1/2, the magnetic flux-density falls off as a power rather than an exponential\cite{babaev2009magnetic}. We believe that this will apply also to the vortices in our model. This, together with the long-range interaction between the vortices, means that they cannot be described as non-interacting anyons even at very large distances, and we refer to them as fractionally charged quasiparticles.  In a broader sense, it is worth noting that although the arbitrary flux fractionalization in multiband superconductors is an effect beyond standard symmetry and topology considerations, yet these objects are remakably stable. Similarly, the quasiparticles we described above is not captured by the  usual topological arguments, yet we expect them to be observable. 

A very interesting situation occurs when there is a quantum phase transition that disorders the $\theta_-$ mode by proliferation of vortices carrying winding only in $\theta_-$, while the $\tilde\theta$ phase is still ordered. In fact, a sequence of phase transitions disordering various combinations of phases as a consequence of proliferation of various vortex combinations was demonstrated in bosonic superfluid systems as a consequence of drag\cite{Dahl2008preemptive,svistunov2015superfluid}. In that case the charge $e/2k$ half vortices will have a finite energy and the CFQH liquid would have a distinct topological order different from the $\nu=1/k$ Laughlin state.
The possibility of such a phase with fractionalized anyons is one of the main results of this paper.

\noindent
\emph{\bf{The effective hydrodynamical action and the edge modes:}} The QH edge modes are most easily understood in the context of a hydrodynamic theory, where the electromagnetic current is related to the hydrodynamic $b_\mu$ field by $J^\mu = \frac 1 {2\pi} \epsilon^{\mu\nu\sigma} \partial_\nu b_\sigma$. In the Lagrangian,
\begin{align} \label{hydact}
\mathcal{L}_{hyd} &= - \frac k {4\pi } \epsilon^{\mu\nu\sigma} b_\mu \partial_\nu b_\sigma - \frac e {2\pi} \epsilon^{\mu\nu\sigma} (\delta A_\mu+\frac{1}{e}C_\mu) \partial_\nu b_\sigma + \half\cdot b_\mu j^\mu+  \frac \mu {8\bar\rho \pi^2} E_b^2 - \frac{2V_f+V_b}{16\pi^2} B_b^2  \nonumber \\
&-  \delta\rho_- \partial_0 \theta_-   -  \frac {\bar\rho}{m_b} (\vec\nabla \theta_-)^2 - \frac{V_b}{4} \delta\rho_-^2  \,  
\end{align}
 
The first line is the Wen-Zee hydrodynamic Lagrangian augmented by  terms derived from the GLCS theory (see e.g. \oncite{hankvo}), and the second line is the gapless $\theta_-$ mode.   $j_\mu$ is the quasiparticle current due to vortices in the $\tilde\theta$ field. From \eqref{hydact} we can, using standard methods, derive the edge action. For an edge along the $x$-direction, the topological term $\sim bdb$ gives the chiral boson action,
\be{edgelag}
{\cal L}_{c} = \frac k {4\pi} \phi (\partial_t - v_f \partial_x )\partial_x \phi - \frac 1 {2\pi} \partial_x\phi (eA_x - C_x)
\ee
which differs from the usual result in that $\phi$ couples not only to the electromagnetic potential $A_x$, but also to the auxiliary potential $C_x$. As usual, $v_f$ is a non-universal drift velocity induced by the edge potential. Thus, there is a charged edge current $J^{em}_x = \frac e {2\pi}\partial_x \phi$, just as for a regular QH liquid, but also, as expected, a neutral counter-flowing bosonic edge mode $J^{bos} = -J^{em}$.

To see the consequences of this, we recall that the thermal current in a one-dimensional channel is given by,
\be{termcur}
J_T = \frac {\pi k_B^2 T^2} {12 v} (c - \bar c)
\ee
where $v$ is the velocity of the $\tilde\theta$ mode, $k_B$ the Boltzmann constant and $c$ and $\bar c$ the central charges of the right- and left-moving modes respectively. In our case, $c= \bar c =1$ so there is no heat current which is an experimentally detectable signature of the counter-flow. We should caution that this conclusion only holds in the ideal system where the $\tilde\theta$ and $\theta_-$ modes are fully decoupled also at the edge. If not, heat could leak from the edge into the gapless $\theta_-$ mode in the bulk.

A simpler system, consisting of a single charged fermion and a single neutral boson, can be analyzed in the same way as above, and there is only one single counter-flow mode related to the phase difference $\theta_f-\theta_b$. Thus, there are no other gapless bulk states that could interfere with heat transport measurements on the edge. In this case, there is a definite prediction that you have charge transport, but no heat transport.

\noindent
\emph{\bf{Summary and outlook:}}
Mixtures of bosonic superfluids and quantum Hall liquids can be described by a multicomponent theory of elementary and composite bosons. Using a model with two neutral bosons and one fermion in a counterflowing, or ``Borromean" state, we identify two phases of what we call a counterflowing quantum Hall liquid.
The first has the same topological order as a Laughlin state, but supports quasiparticles with arbitrary fractional electric charge and a long-range logarithmic interaction. In analogy with ``unquantized" vortices in superconductors \cite{Iguchi2023,zheng2024observation,Zhou2024}, we expect these quasiparticles to be observable. Note that fractional vortices in superconductors become stable near boundaries even when being linearly confined in bulk\cite{silaev2011stable,maiani2022vortex}. In our case linear confining interactions would arise if the symmetry associated with $\theta _-$ is explicitly broken down to a discrete symmetry.
The other phase, again with the same Hall resistance as a Laughlin liquid, but with unconfined anyons with fractional charge and statistics, quantized to half of the Laughlin values, thus defining a distinct topological order. We also analyzed the edge properties of the counterflowing quantum Hall liquid.

The counterflowing quantum Hall liquid could be realized in Bose-Fermi mixtures of ultracold atoms \cite{bloch2008many,bloch2012quantum,lewenstein2012ultracold},
where great progress was made recently in experimentally realizing bosonic counterflow state \cite{zheng_pan2025counterflow}. However, they may potentially also be realized in condensed matter systems with multiple bands or layers, where some contain unpaired electrons, and others bosons, which can be either excitons or charged Cooper pairs.
An example of a different kind of three-component counterflow state in an electronic system was reported in a multiband system \cite{Grinenko2021state}.
Especially interesting are layered moire systems, which have become increasingly promising platforms for multicomponent electronic systems, and which could be used to realize various Hubbard-type models\cite{gu2022dipolar}.

Clearly, the analysis in this paper can be extended to a large new class of quantum liquids by considering a higher number of bosons, and also other quantum Hall states in the Abelian hierarchy.

\noi
{\bf {Acknowledgments}}\\
THH thanks Steve Simon for useful comments, and 
JXH thanks Qingdong Jiang and Gabriel Cardoso for helpful discussions. EB thanks Boris Svistunov for discussions,
EB was supported by the Knut and Alice Wallenberg Foundation project grant and via the Wallenberg Center for Quantum Technology (WACQT), the Wallenberg Initiative Materials Science for Sustainability (WISE), and the Swedish Research Council on Grant 2022-04763.

\appendix

\section { Supplementary Material }
\subsection{General three-component model}
For the general case $m_1\neq m_2 \neq m_f$, and after implementing the Mott constraint, matter part of the action \eqref{xx} can be written as
\be{eq:tilderepMott}\label{xx}
S &=& \int d^3x\, -\delta\rho_f(\partial_0\tilde{\theta}-\delta a_0) -
\delta\rho_-\partial_0\theta_- - \tilde{\gamma}(\vec\grad\tilde\theta-\delta\vec{a})^2 - \gamma_-(\vec\grad\theta_-)^2 \nonumber \\
  &-& 2\gamma_d(\vec\grad\tilde\theta-2\delta\vec{a})\cdot(\vec\grad\theta_-) - V(\delta\rho_f,\delta\rho_1,\delta\rho_2)
\ee
Where we kept the quadratic terms only, and neglected the density gradient term. The constants are given by
\be{eq:gammas}
\tilde\gamma = \frac{(\gamma_1+\gamma_2)\gamma_f}{2(\gamma_1+\gamma_2+\gamma_f)},\,\gamma_- = \frac{(\gamma_1+\gamma_2)\gamma_f+4\gamma_1\gamma_2}{2(\gamma_1+\gamma_2+\gamma_f)},\,\gamma_d = \frac{(-\gamma_1+\gamma_2)\gamma_f}{2(\gamma_1+\gamma_2+\gamma_f)} \, .
\ee

Where $\gamma_\alpha=\bar\rho/m_\alpha$ for $\alpha=f,1,2$. Note that there is a mixed gradient term $\sim \gamma_d$, which couples the $\tilde\theta$ and $\theta_-$ fields. 

In the special case where $m_1=m_2=m_b$ we have $\gamma_d=0$, and with $\mu=m_f+\half  m_b$, $\tilde\gamma=\bar\rho/4\mu,\, \gamma_-=\bar\rho/2m_b$,  the action becomes
\be{eq:m1=2action}
\nonumber S &=& \int d^3x\,  -\delta \rho_f (\partial_0\tilde\theta-\delta a_0)  - \delta\rho_- \partial_0 \theta_- - \frac {\bar\rho}{2\mu} (\vec\grad \tilde\theta - \delta\vec a)^2 - \frac {\bar\rho}{m_b} (\vec\grad \theta_-)^2 \\
            &-& V(\delta\rho_f,\delta\rho_1,\delta\rho_2)
\ee
Further specializing to  $m_f=m_1=m_2=m$, we retain the action \eqref{eq:m1=2action} in the main text. 

The expression \eqref{xx} given in the main text does not have any  mixed gradient terms, but at the expense of the two modes $\tilde\theta$ and $\theta_-$ have in cross terms in the symplectic form, and thus off-diagonal commutation relations. We now show that there is a choice of basis that completely decouples the two modes both in the Hamiltonian, and in the symplectic form. 

By making a $SO(2)$ rotation to the $(\delta\rho_f,\tilde\theta)$ and $(\delta\rho_-,\theta_-)$ modes with the matrix
 \be{eq:O2matrix}
    O^{(2)} = \begin{pmatrix}
             \cos\frac{\alpha}{2} & \sin\frac{\alpha}{2} \\
             \sin\frac{\alpha}{2} & -\cos\frac{\alpha}{2}
              \end{pmatrix}
    \ee
where $\bar\gamma=(\tilde\gamma+\gamma_-)/2,\delta\gamma=(\tilde\gamma-\gamma_-)/2,\tan\alpha=\gamma_d/\delta\gamma$,  
   
    i.e. by defining $(\delta\varrho_+,\vartheta_+)=\cos\frac{\alpha}{2}(\delta\rho_f,\tilde\theta)+\sin\frac{\alpha}{2}(\delta\rho_-,\theta_-),\, (\delta\varrho_-,\vartheta_-)=\sin\frac{\alpha}{2}(\delta\rho_f,\tilde\theta)-\cos\frac{\alpha}{2}(\delta\rho_-,\theta_-)$, we rewrite the action as
    
    \be{eq:O2dragelimination}
    S &=& \int d^3x\, -\delta\varrho_+(\partial_0\vartheta_+ - \cos\frac{\alpha}{2}\delta a_0) - \delta\varrho_-(\partial_0\vartheta_- - \sin\frac{\alpha}{2}\delta a_0) \nonumber \\
      &-& (\bar\gamma+\sqrt{\delta\gamma^2+\gamma_d^2})(\vec{\nabla}\vartheta_+ - \cos\frac{\alpha}{2}\delta\vec{a})^2 - (\bar\gamma-\sqrt{\delta\gamma^2+\gamma_d^2})(\vec{\nabla}\vartheta_- - \sin\frac{\alpha}{2}\delta\vec{a})^2 \nonumber \\
      &-& V(\delta\rho_f,\delta\rho_1,\delta\rho_2)
    \ee
we obtained the action in terms of two decoupled, although charged (and thus interacting electromagnetically), modes as claimed in the main text.

\subsection{Basis independence of vortex energies}
In this section we show by an explicit example that our conclusions about vortex fractionalization does not rely on the particular choice of fields representation $\tilde\theta$ and $\theta_-$.

A possible alternative choice of basis is, 
\be{eq:varthetaAction}
S &=& \int d^3x\, -\delta\rho_1(\partial_0\vartheta_1+\delta a_0) - \delta\rho_2(\partial_0\vartheta_2+\delta a_0) \\ \nonumber
    &-& \frac{1+m_f/m_b}{2}\frac{\bar\rho}{2\mu}[(\nabla\vartheta_1+\delta\vec{a})^2+(\nabla\vartheta_2+\delta\vec{a})^2] + \frac{m_f}{m_b}\frac{\bar\rho}{2\mu}(\nabla\vartheta_1+\delta\vec{a})\cdot(\nabla\vartheta_2+
    \delta\vec{a}) \\ \nonumber
    &-& V(\delta\rho_f,\delta\rho_1,\delta\rho_2) 
\ee
and hence
\be{eq:varthetaHamiltonian}
H = \int d^2x\, \frac{1+m_f/m_b}{2}\frac{\bar\rho}{2\mu}[(\nabla\vartheta_1+\delta\vec{a})^2+(\nabla\vartheta_2+\delta\vec{a})^2] - \frac{m_f}{m_b}\frac{\bar\rho}{2\mu}(\nabla\vartheta_1+\delta\vec{a})\cdot(\nabla\vartheta_2+\delta\vec{a})
\ee
where $\vartheta_{1,2}=\theta_{1,2}-\theta_f$, and we negelected the density fluctuation terms. Now a $(1,0,0)$ vortex corresponds to $\vec\grad\vartheta_1=-\frac{1}{r}\hat{e}_{\phi}, \, \vec\grad\vartheta_2=-\frac{1}{r}\hat{e}_{\phi}$; a $(0,1,0)$ vortex corresponds to$\vec\grad\vartheta_1=\frac{1}{r}\hat{e}_{\phi}, \, \vec\grad\vartheta_2=0$; a $(0,0,1)$ vortex corresponds to $\vec\grad\vartheta_1=0, \, \vec\grad\vartheta_2=\frac{1}{r}\hat{e}_{\phi}$. With straightforward calculations, we see that by minimizing energy density with variational Chern-Simons flux $\delta\vec{a}=\frac{\Phi}{2\pi}\frac{1}{r}\hat{e}_\phi$, we obtain stabilized vortex solution which is independent of whether we choose $(\tilde\theta,\theta_-)$ or $(\vartheta_1,\vartheta_2)$ representation.

As a matter of fact, the vortex property is independent of basis choice. If we treat the Hamiltonian as a functional of $(\theta_f,\theta_1,\theta_2,\delta\vec{a})$ fields, given the winding in any one of the phase field and flux parametrized by $\Phi$, the Hamiltonian is simply a function of $\Phi$, independent of choice of variables for phase fields. Hence the energy-density-minimization procedure is invariant for different representations, and so the result of it.
 
%

\end{document}